%Paper: hep-ph/9504292
%From: dudas@amoco.saclay.cea.fr (Emilian Dudas)
%Date: Wed, 12 Apr 1995 15:54:51 +0200
%Date (revised): Wed, 12 Apr 1995 23:10:54 +0200

\magnification=1095
\baselineskip=15pt
\hskip 10cm
SPhT Saclay T95/027
\vskip 2pt
\hskip 10cm
MPI-PTh 95-33
\vskip 2pt
%\hskip 10cm
%hep-ph/9504292
%\vskip 2pt
\hskip 10cm
March 1995
\vskip 1cm
\centerline{{\bf YUKAWA MATRICES FROM A SPONTANEOUSLY BROKEN ABELIAN SYMMETRY}}
%\vskip 6pt{\bf\ }
%\centerline{{\bf BROKEN ABELIAN SYMMETRY}}
\vskip 24pt
\centerline{{\bf E. Dudas}}
\vskip 8pt
\centerline{CEA, Service de Physique Th\'eorique, CE-Saclay}
\centerline{F-91191 Gif-sur-Yvette Cedex, FRANCE}
\vskip 12pt
\centerline{{\bf S. Pokorski}\footnote{$ ^\ast $}{On leave from the Institute
of Physics, Warsaw. Supported in part by the Polish Committee for Scientific
Research and by the EC grant Flavourdynamics}}
\vskip 8pt
\centerline{Max-Planck Institut f\"ur Physik Werner-Heisenberg-Institut}
\centerline{F\"ohringer Ring 6, D-80805 M\"unchen, GERMANY}
\vskip 12pt
\centerline{{\bf C.A. Savoy}}
\vskip 8pt
\centerline{CEA, Service de Physique Th\'eorique, CE-Saclay}
\centerline{F-91191 Gif-sur-Yvette Cedex, FRANCE}
\vglue 1.2truecm
\centerline{{\bf ABSTRACT}}
\vskip 8pt

We classify all the phenomenologically viable fermion mass matrices
coming from a spontaneously broken abelian symmetry $ U(1)_X, $ with one and
two
additional chiral fields of opposite charges $ X=\pm 1. $  We find that the
non-trivial K\"ahler metric can fill zeroes of
the fermion mass matrices up to phenomenologically interesting values. A
general
anomaly analysis shows that for one additional chiral field the only way to
achieve anomaly cancellation is by use of the Green-Schwarz mechanism. For
two additional fields with $ X=\pm 1 $ and negative charge differences in the
lepton
sector the anomalies can however be directly put to zero. This case gives a
unique prediction for the ratio of the two Higgs scalars of MSSM, $ {\rm tg} \
\beta \sim { m_t \over m_b} (sin {\theta_c})^ 2$, where $\theta_c$ is the
Cabibbo angle.
\vglue 1.5truecm
{\sl Submitted to Physics Letters B\/}
\pageno=0
\vfill
\eject

{\bf 1.\nobreak\ }The Froggat-Nielsen [1] idea of understanding fermion mass
textures in terms of
additional (to the Standard Model) $ U(1) $ symmetries (global or gauged) has
recently been reviewed in the context of supersymmetric models
[2], [3], [4], [6]. The recent investigations go into
two directions. First, several
concrete models for the mass textures have been proposed
[5]. Second, a
potential connection has been pointed out between the mechanism for the
fermion mass generation and the Green-Schwarz mechanism for anomaly
cancellation, in case of gauged $ U(1) $ symmetries [7]. This
connection has been
explored in some detail in the framework of models with \lq\lq stringy\rq\rq\
$ U(1) $
symmetries spontaneously broken slightly below the string scale
[8],[9] (another way of obtaining Froggatt-Nielsen structures in effective
superstrings is described in [18]).

In this paper we generalize and systematize these results in several aspects.
We begin with a general search for phenomenologically acceptable fermion mass
textures in a model with one additional $ U(1) $ symmetry and one or two
additional chiral fields of opposite charge (singlets with respect to the SM
gauge group but
charged under the new $ U(1)$).  To make our search effective we derive general
formulae for the CKM matrix elements in terms of the entries in the mass
matrices (the formulae are given in the Appendix) which provide a useful
supplement to the already existing expressions in the literature  for the mass
eigenvalues [4].
Equipped with both, we are able to find in our model all
phenomenologically acceptable mass textures. It is
interesting to observe that the number of acceptable textures is strongly
limited: to four if we insist on getting all the physical observables with
precisely the right order of magnitude and to few more if we allow for order $
\lambda $
(Cabbibo angle) deviations in some of them. One can argue that this
additional freedom is acceptable, given the fact that the model can predict
the mass matrix entries only up to coefficients of order $ O(1).$
The textures
with some of the entries being zero are possible only if we accept to work
with $ O(\lambda) $ precision for some of the observables. It is interesting
to notice
that in such cases, in general, the zeroes disappear after renormalization
which brings the K\"ahler potential to the canonical form. Thus, the final
results for the mass matrices depend on those renormalization effects
and do not have any zeroes.

The second part of the paper is devoted to a discussion of the potential
anomalies introduced by the additional $ U(1) $ symmetry. Starting with the
observation that the quark mass matrices are invariant under shifts in $ U(1)
$
charges, it is legitimate to ask if this freedom can be used to cancel the
anomalies. We derive a general formula for the anomalies in terms of the
quark and lepton masses and prove that anomaly cancellation is
impossible for physically acceptable textures with one singlet, or with two
singlets if all the charge differences
which enter the mass textures are positive. However, the Green-Schwarz
mechanism of anomaly cancellation can always be realized (by the mentioned
above redefinition of the $ U(1) $ fermion charges) and, in fact, this
requirement gives no constraint on the textures. Finally, we show that,
with two singlets and negative charge differences, one finds solutions
with anomaly cancellation achieved in the
standard way, without asking for the Green-Schwarz mechanism. In this case,
the prediction of the model is $ {\rm tg} \ \beta \sim{ m_t \over m_b}\lambda^
2\sim 2, $ independent of the possible
$U(1)_X$ charge assignment for leptons.

\vskip 24pt

{\bf 2.}\nobreak\ The physical quantities, the fermion masses and the
Kobayashi-Maskawa matrix
are obtained by diagonalization of the mass matrices

$$ \eqalign{ U_Lm^UU^+_R & = {\rm diag} \left(m_u,m_c,m_t \right) \cr
D_Lm^DD^+_R & = {\rm diag} \left(m_d,m_s,m_b \right) \cr L_Lm^LL^+_R & = {\rm
diag} \left(m_e,m_\mu ,m_\tau \right)\ , \cr} \eqno (1) $$
the CKM matrix being given by $ V_{ {\rm CKM}} = U_LD^+_L. $ There are 9
masses, 3 mixing angles
and 1\nobreak\ CP violating phase to be compared with the experimental data.
 The known experimental data has a strongly hierarchical
structure in the three families. Using the Cabibbo angle $ (\lambda  = {\rm
sin} \ \theta_ c\sim 0.22) $ as a small expansion parameter,
the order of magnitude values of the mass ratios  can be written in the form

$$ {m_u \over m_t} \sim  \lambda^ 8,\ \ {m_c \over m_t} \sim  \lambda^ 4,\ \
{m_d \over m_b} \sim  \lambda^ 4,\ \ {m_s \over m_b} \sim  \lambda^ 2,\ \ {m_e
\over m_\tau}  \sim  \lambda^ 4,\ \ {m_\mu \over m_\tau}  \sim  \lambda^ 2
\eqno (2) $$
at a high energy scale $ M_{ {\rm GUT}} \sim  10^{16} {\rm GeV} . $ The
hierarchical structure is also transparent in the CKM matrix. In the
Wolfenstein parametrization of the known experimental data, $ V_{ {\rm CKM}} $
writes as
$$ V_{ {\rm CKM}} = \left( \matrix{ 1 - {\lambda^ 2 \over 2}  & \lambda  &
\lambda^ 3 A \chi (\rho +i\eta) \cr -\lambda  &  1-{\lambda^ 2 \over 2}  &
\lambda^
2A \chi \cr \lambda^ 3A \chi (1-\rho +i\eta)  &  -\lambda^ 2A \chi  & 1 \cr}
\right)\ , \eqno
(3) $$
where $ A\sim 1 $, $ \lambda <\rho ,\eta  < 1$ and the factor
$\chi \simeq 0.7$ describes the running from $M_Z$ to $M_{GUT}$ [10]. The CP
phase is neglected in this paper, so for our purposes $ V_{ {\rm CKM}} $ is
just an
orthogonal matrix.

A simple way to understand these structures is to postulate a family
(horizontal) gauge symmetry spontaneously broken by the vacuum expectation
values (vev's) of some scalar fields $\phi$ which are singlets under the
Standard Model  gauge group. The hierarchy of fermion masses and mixing
angles is then explained by assignement of charges of the horizontal
group such that invariant terms in the lagrangian (or superpotential in the
supersymmetric case) have the form $({<\phi> \over M})^{n_{ij}} \bar \psi_i
\psi_j H$ (after decoupling of the heavy fields), where $\psi_i$ are the
SM fermions, $H$  is a Higgs field and $M$ is a large scale.
Postulating $ \varepsilon  \equiv  {\langle
\phi\rangle \over M} \simeq \lambda$ (the Cabibbo angle) one can easily
explain hierarchies in the effective Yukawa couplings, even in the simplest
case of abelian $U(1)_X$ symmetry, with all the coefficients of the higher
dimension operators of the order $O(1)$.

\vskip 24pt

{\bf 3.}\nobreak\ In the following we investigate systematically all
phenomenologically acceptable Yukawa matrices which can be obtained
in a model with one horizontal $U(1)_X$ symmetry and

a) one additional SM gauge singlet $ \phi $ of charge $ X=-1 $

b) two additional SM singlets $ \phi $ and $ \tilde \phi $ of opposite
charges $ X=-1 $ and $ \tilde X=+1 $ respectively.

To simplify the task it is very useful to have explicit formulae for
the mass eigenvalues and the CKM mixing angles (i.e. for the rotation
matrices $U_L$ and $D_L$) directly in terms of the original Yukawa
matrix entries.
The former exist in the literature whereas the later are given in [11] under
the assumption of hierarchy between all rows and all columns,
as in the original proposal of Froggatt-Nielsen [1].
In this case $ U_L $ and $ D_L $
are almost diagonal and the small mixing angles can be analytically computed.
However, this assumption is too restrictive for a systematic study of all
phenomenologically acceptable mass matrices in the model considered.
Therefore, we present in the Appendix the formulae for the CKM mixing
angles derived under a weaker assumption,
namely that
$ m_{33} \geq  m_{ij}, $ $(i,j) \not=  (3,3) $ for all the fermion matrices
and that the rotation matrices $U_L$ and $D_L$ consist of small rotations,
at most of order $O(\lambda)$. The latter assumption follows from the
naturalness argument: the smallnes of the CKM rotations should not be due
to a relative fine-tuning of the $U_L$ and $D_L$.

The derivation is based on the formalism developed in [11]. The matrices $ m $
can be diagonalized by three successive rotations in
the
$ (2,3), $ $ (1,3) $ and $ (1,2) $ sectors (with the angles $ S_{23},
S_{13} $ and $ S_{12}$),

$$ Q_L = \left( \matrix{ 1  & -S^Q_{12}  & 0 \cr S^Q_{12}  & 1  & 0 \cr 0  & 0
 & 1 \cr} \right) \left( \matrix{ 1  & 0  & -S^Q_{13} \cr 0  & 1  & 0 \cr
S^Q_{13}  & 0  & 1 \cr} \right) \left( \matrix{ 1  & 0  & 0 \cr 0  & 1  &
-S^Q_{23} \cr 0  & S^Q_{23}  & 1 \cr} \right)\ , \eqno (4)
$$
where  $ (Q_L=U_L,D_L) $.
For all the textures to be discussed in the following, the matrices $ U_R $
and $ D_R $
(with the angles $ S^{\prime Q}_{23}, $ $ S^{\prime Q}_{13} $ and $ S^{\prime
Q}_{12}) $ are almost diagonal in the $ (2,3) $ and $ (1,3) $
sectors. In the $ (1,2) $ sector they can be either almost diagonal or almost
antidiagonal, the physical CKM matrix being the same in the two cases.
Without loss of generality, we can consider the almost diagonal
case; the antidiagonal one differs by just a permutation of it and need not a
particular treatment.

We consider now case (a), with one SM gauge singlet.
Due to our ignorance of the
exact coefficients multiplying the powers of $ \varepsilon $ in the Yukawa
couplings and of
the present experimental uncertainties of (2) and (3) we allow for mass
matrices resulting in predictions compatible with these data up to few
deviations $ O(\lambda) $ in some observables. Our goal is to classify the
best textures according to the number of these deviations.
 Supersymmetry is assumed, so that the
Yukawa couplings are encoded in the superpotential.
 In this case, if the sum of the $ U(1)_X $ charges of the fields
corresponding
to a specific Yukawa interaction is negative, the holomorphicity of the
superpotential forbids this particular Yukawa coupling and the corresponding
element is zero. Zeroes in the mass matrices are generically
indicating the appearance of additional symmetries beyond the Standard Model
and were largely investigated in the literature Refs.[12],
[13], [14].

 The part of the  $U(1)_X$ invariant superpotential of the theory
responsible for the quark and lepton masses is

$$ \eqalignno{ W & = \sum^{ }_{ ij} \left[Y^U_{ij}\theta \left(q_i+u_j+h_2
\right) \left({\phi \over M} \right)^{q_i+u_j+h_2} Q^i U^j H_2 \right. &  \cr
 &  +
Y^D_{ij}\theta \left(q_i+d_j+h_1 \right) \left({\phi \over M}
\right)^{q_i+d_j+h_1}Q^iD^jH_1 &  \cr  & \left.+Y^E_{ij}\theta
\left(l_i+e_j+h_1 \right) \left({\phi \over M} \right)^{l_i+e_j+h_1}L^iE^jH_1
\right]\ , & (5) \cr} $$
where $ Y^{U,D,E}_{ij} $ are numbers of $ O(1) $.
We denote the fields and their $ U(1)_X $ charges by the same
capital and small letters, respectively.

The general K\"ahler potential
consistent with the $ U(1)_X $ symmetry reads

$$ K = \sum^{ }_{ \Phi =Q^i,U^i,D^i,L^i,E^i,H_1,H_2}Z^\Phi_{ ij}
\left[\theta \left(\varphi_ i-\varphi_ j \right) \left({\phi \over M}
\right)^{\varphi_ i-\varphi_ j}+\theta \left(\varphi_ j-\varphi_ i \right)
\left({\phi^+ \over M} \right)^{\varphi_ j-\varphi_ i}
\right]\Phi^ i\Phi^{+j}\ , \eqno (6) $$
where $Z^\Phi_{ij}$ are numbers.
The physical Yukawa couplings are obtained by the canonical normalization
of the kinetic terms and are given by
$$ Y^U_{ij} = \left(K^{-1/2} \right)^{i\prime}_ i {\partial^ 3W \over \partial
Q^{i\prime} \partial U^{j\prime} \partial H_2} \left(K^{-1/2}
\right)^{j\prime}_ j\ , \eqno (7) $$
and similar expressions for $ Y^{D,E}. $ The potential effect of the kinetic
terms in (7) is to remove
the eventual zeroes. Consequently, they can change the physical
predictions of the texture. Some examples with zero filling of
phenomenological interest will be given below.

 Using the shorthand notation $ \hat \varepsilon^
q=\theta( q)\varepsilon^ q $ and defining the parameter $x$ by
$ h_1=-q_3-d_3+x $ , we can write
$$ Y^D = \varepsilon^ x \left( \matrix{\hat \varepsilon^{ q_{13}+d_{13}}  &
\hat \varepsilon^{ q_{13}+d_{23}}  & \hat \varepsilon^{ q_{13}} \cr\hat
\varepsilon^{ q_{23}+d_{13}}  & \hat \varepsilon^{ q_{23}+d_{23}}  & \hat
\varepsilon^{ q_{23}} \cr\hat \varepsilon^{ d_{13}}  & \hat \varepsilon^{
d_{23}}  & 1 \cr} \right)\ , \eqno (8) $$
where $ q_{ij} = q_i-q_j, $ $ d_{ij}=d_i-d_j. $ Similar expressions hold for $
Y^E $ (with the same $ x $ since
$ Y^D_{33} \sim Y^E_{33} $, see ref. [15], implying
 $ h_1=-l_3-e_3+x $)  and for $ Y^U $ (with $
x=0 $ in order to accomodate a heavy top mass).
Only the charge differences appear in (8), so a specific texture will fix
them. The residual freedom in the values of the charges will be further
restricted by the anomaly cancellation conditions, to be discussed later on.

Let us now turn to the allowed textures, according to the announced
$O(\lambda)$ deviation rule. An important point to emphasize is that a
permutation
of the first two columns independently for $Y_U$, $Y_D$ (first two lines
simultaneously for $Y_U$, $Y_D$) of the mass matrices has as only effect
changing the right-handed (left-handed) diagonalizing angles $ S^{\prime}_{
ij} \left(S_{ij} \right), $ the
masses and $ V_{ {\rm CKM}} $ being unchanged.
So all proposed solutions have 7 other possibilities, related to each other by
the permutations $ u_1 \longleftrightarrow u_2, $ $ d_1
\longleftrightarrow d_2, $ $ q_1 \longleftrightarrow q_2. $

To search for all phenomenologically acceptable mass matrices we scan
over all possible charge assignements and use the formulae in the
Appendix to check the resulting matrices.

The results are:

- Only one solution which gives the right order of
magnitude in $\lambda$ for all masses and mixing angles and it is
given by

$$ q_{13}=3,\ \ q_{23}=2,\ \ u_{13}=5,\ \ u_{23}=2,\ \ d_{13}=1,\ \ d_{23}=0
\eqno (9) $$
and corresponds to the original proposal of Ref.[1]. It is
characterized by having
no zeroes in the mass matrices. Four other solutions with one
$O(\lambda)$  deviation
are obtained by
the changes $ u_{i3} \longrightarrow u_{i3}\pm 1, $ $ d_{i3} \longrightarrow
d_{i3}\pm 1, $ $ i=1,2, $ and others with two deviations in
combining them.

\vskip 5pt

- One solution with two $O(\lambda)$ deviations corresponds to one zero (filled
as in
the Eq.(7)) in $ Y^D. $ The charge assignments for $ q_{ij}, $ $ u_{ij} $ and $
d_{ij} $ are
$$ q_{13}=4,\ \ q_{23}=3,\ \ u_{13}=4,\ \ u_{23}=1,\ \ d_{13}=1,\ \ d_{23}=-1\
. \eqno (10) $$
The two deviations are $ {m_d \over m_b} \sim  \lambda^ 5 $ and $ V_{cb} \sim
\lambda^ 3. $

\vskip 5pt

- One solution with two $O(\lambda)$ deviations, with two filled zeroes in $
Y^D. $
The
charge assignements are
$$ q_{13}=4,\ \ q_{23}=3,\ \ u_{13}=4,\ \ u_{23}=1,\ \ d_{13}=d_{23}=-1\ .
\eqno (11) $$
The deviations are $ {m_d \over m_b} \sim  \lambda^ 3 $ and $ V_{cb} \sim
\lambda^ 3. $

In the last two solutions, the zero filling affects only the right angles $
S^{\prime}_{ 13} $
and $ S^{\prime}_{ 23}, $ with a possible effect in the analysis of the
neutral currents constraints [16], [17].

\vskip 5pt

- One solution with two $O(\lambda)$ deviations with two filled zeroes in $ Y^U
$ and
two
filled zeroes in $ Y^D. $ The charge assignment is
$$ q_{13}=-2,\ \ q_{23}=-3,\ \ u_{13}=10,\ \ u_{23}=7,\ \ d_{13}=6,\ \
d_{23}=5\ . \eqno (12) $$
The two deviations are in $ V_{ub} $ and $ V_{cb}. $ Before the zero filling $
V_{ub}\sim \lambda^{ 12} $ and $ V_{cb}\sim \lambda^{ 11} $
so the example would not fulfill our criterion of $O(\lambda)$ deviation, the
predictions being completely wrong. After the zeroes filling, the predictions
change $ V_{ub} \sim  \lambda^ 2, $ $ V_{cb} \sim  \lambda^ 3 $ and are much
closer to the correct results, Eq.(3).
This example shown that the kinetic terms in Eq.(7) can
play an important role and cannot generally be neglected.

The charges of the fields for the textures which results from the most
successful
solution (9) are collected in Table 2.
\vskip 10pt

Case (b), with a vector-like pair of singlets, offers (in addition to (9) which
is
a solution in this case, too) additional textures
which fit exactly the experimental
data (2) and (3). They are characterized by a big, negative charge difference,
all the others being positive. For the up-quarks, the assignement is
$$ u_{13}=-11,\ \ u_{23}=2,\ \ q_{13}=3,\ \ q_{23}=2\ . \eqno (13) $$
For the down-quark masses it is

$$ d_{13}+2x = -7,\ \ d_{23}=0,\ \ q_{13}=3,\ \ q_{23}=2, \eqno (14) $$
where $ \varepsilon^ x = {\lambda_ b \over \lambda_ t}. $ This corresponds to
the matrix
$$ Y_D = \varepsilon^ x \left( \matrix{ \varepsilon^ 4  & \varepsilon^ 3  &
\varepsilon^ 3 \cr \varepsilon^ 5  & \varepsilon^ 2  & \varepsilon^ 2 \cr
\varepsilon^ 7  & 1  & 1 \cr} \right)\ . \eqno (15) $$
The two assignements for $u_{13}$ and $u_{23}$ in (9) and (13) can be combined
with those for $d_{13}$ and $d_{23}$ in (9) and (14) to give four solutions.
Remark that the textures with
negative charge differences are characterized by an anti-hierarchy structure in
the
first column, the two other columns being the same as in Eq.(9). There are
other
solutions with some $O(\lambda)$ deviations the case (b) which are not
displayed here. They are less interesting than the corresponding one in case
(a) in the sense that the K\"ahler potential plays no role in their structure.
All of the above solutions are collected in Table 1.

A universal prediction of all these models is $V_{us} \sim V_{ub}/V_{cb}$.
On the other hand, the parameter $x$, related to the $tg \beta$ parameter
of the MSSM by $tg \beta = {m_t \over m_b} {\lambda}^x$ is arbitrary, being
apriori unconstrained by the experimental data (2),(3).
Finally, we use the assignements (9)-(14) in the analysis of the gauge anomaly
cancellation conditions to be discussed in the next paragraph.
\vskip 24pt

$$ \vbox{\offinterlineskip\halign{
& \vrule#& \strut\kern.3em# \kern0pt
& \vrule#& \strut\kern.3em# \kern0pt
& \vrule#& \strut\kern.3em# \kern0pt
& \vrule#& \strut\kern.3em# \kern0pt
&\vrule#\cr
\noalign{\hrule}
height2pt& \omit&& \omit&& \omit&& \omit&\cr
&$\displaystyle \hfill \matrix{ {\rm Number} \cr {\rm singlets} \cr}
\hfill$&&$\displaystyle \hfill Y_U \hfill$&&$\displaystyle \hfill Y_D
\hfill$&&$\displaystyle \hfill \matrix{ {\rm Number} \cr {\rm deviations} \cr}
\hfill$&\cr
height2pt& \omit&& \omit&& \omit&& \omit&\cr
\noalign{\hrule}
height8pt& \omit&& \omit&& \omit&& \omit&\cr
&$\displaystyle \hfill 1 \hfill$&&$\displaystyle \hfill \left( \matrix{
\lambda^ 8  &    & \lambda^ 5  &    & \lambda^ 3 \cr \lambda^ 7  &    &
\lambda^ 4  &    & \lambda^ 2 \cr \lambda^ 5  &    & \lambda^ 2  &    & 1 \cr}
\right) \hfill$&&$\displaystyle \hfill \left( \matrix{ \lambda^ 4  &    &
\lambda^ 3  &    & \lambda^ 3 \cr \lambda^ 3  &    & \lambda^ 2  &    &
\lambda^ 2 \cr \lambda  &    &  1  &    & 1 \cr} \right)
\hfill$&&$\displaystyle \hfill 0 \hfill$&\cr
height8pt& \omit&& \omit&& \omit&& \omit&\cr
&$\displaystyle \hfill 2 \hfill$&&$\displaystyle \hfill \left( \matrix{
\lambda^ 8  &    & \lambda^ 5  &    & \lambda^ 3 \cr \lambda^ 9  &    &
\lambda^ 4  &    & \lambda^ 2 \cr \lambda^{ 11}  &    & \lambda^ 2  &    & 1
\cr} \right) \hfill$&&$\displaystyle \hfill \left( \matrix{ \lambda^ 4  &    &
\lambda^ 3  &    & \lambda^ 3 \cr \lambda^ 5  &    & \lambda^ 2  &    &
\lambda^ 2 \cr \lambda^ 7  &    & 1  &    & 1 \cr} \right)
\hfill$&&$\displaystyle \hfill 0 \hfill$&\cr
height8pt& \omit&& \omit&& \omit&& \omit&\cr
\noalign{\hrule}
height8pt& \omit&& \omit&& \omit&& \omit&\cr
&$\displaystyle \hfill 1 \hfill$&&$\displaystyle \hfill \left( \matrix{
\lambda^ 8  &    & \lambda^ 5  &    & \lambda^ 4 \cr \lambda^ 7  &    &
\lambda^ 4  &    & \lambda^ 3 \cr \lambda^ 4  &    & \lambda  &    &  1 \cr}
\right) \hfill$&&$\displaystyle \hfill \matrix{ \left( \matrix{ \lambda^ 5  &
  & \lambda^ 3  &    & \lambda^ 4 \cr \lambda^ 4  &    & \lambda^ 2  &    &
\lambda^ 3 \cr \lambda  &    & {\underline{\lambda}}  &    &  1 \cr} \right)
\cr \cr \cr
\left( \matrix{ \lambda^ 3  &    & \lambda^ 3  &    & \lambda^ 4 \cr \lambda^
2  &    & \lambda^ 2  &    & \lambda^ 3 \cr {\underline{\lambda}}  &    &
{\underline
{\lambda}}  &    &  1 \cr} \right) \cr} \hfill$&&$\displaystyle \hfill 2
\hfill$&\cr
height8pt& \omit&& \omit&& \omit&& \omit&\cr
\noalign{\hrule}
height8pt& \omit&& \omit&& \omit&& \omit&\cr
&$\displaystyle \hfill 1 \hfill$&&$\displaystyle \hfill \left( \matrix{
\lambda^ 8  &    & \lambda^ 5  &    & {\underline{\lambda}}^ 2 \cr \lambda^ 7
&
& \lambda^ 4  &    & {\underline{\lambda}}^ 3 \cr \lambda^{ 10}  &    &
\lambda^ 7
&    & 1 \cr} \right) \hfill$&&$\displaystyle \hfill \left( \matrix{ \lambda^
4  &    & \lambda^ 3  &    & {\underline{\lambda}}^ 2 \cr \lambda^ 3  &    &
\lambda^ 2  &    & {\underline{\lambda}}^ 3 \cr \lambda^ 6  &    & \lambda^ 5
&
& 1 \cr} \right) \hfill$&&$\displaystyle \hfill 2 \hfill$&\cr
height8pt& \omit&& \omit&& \omit&& \omit&\cr
\noalign{\hrule}}}
 $$

\vskip 24pt

{\bf TABLE 1:} Phenomenologically interesting quark mass matrices from a
horizontal
$ U(1)_X $ symmetry. The underlined entries are filled zeroes as in Eq.(7). The
solutions $ \left(Y_U,Y_D \right) $ not separated by a horizontal line can be
freely combined.
The first two columns (independently for $ Y_U,Y_D) $ or the first two lines
(simultaneously for $ Y_U,Y_D) $ can be interchanged without changing the
physical
predictions. Solutions with two singlets and some $ O(\lambda ) $ deviations
are not
displayed.

{\bf 4.}\nobreak\ Let us now turn to the question of anomaly cancellation. The
new abelian gauge
group $ U(1)_X $ is potentially anomalous. It generates, through triangle
diagrams,
mixed anomalies with the standard model gauge group. The
anomaly conditions in connection
with mass textures were considered in [3] and [4] in the particular
case of $ h_1+h_2=0 .$ Moreover, ref.[3]
concentrates on left-right symmetric mass matrices, while
ref.[4] assumes
$ h_1=h_2=0, $ and positive charge differences $ q_{i3}, $ $ u_{i3}, $ $
d_{i3}, $ $ l_{i3}, $ $ e_{i3} $ for $ i=1,2. $

We relax here these restrictions and analyse the general case. Indeed, as
remarked here above there exist acceptable textures with negative $ X $-charge
differences with two singlets. Moreover, the value of $ h_1+h_2 $ constrains
the $ \mu $-term of the MSSM
and there is no compelling reason to assume it to vanish.
The mixed anomalies to be considered are the following
$$ \eqalignno{[ {\rm SU} (3)]^2 {\rm U}(1)_X: & \ \ \ \ \ \ A_3= \sum^ 3_{i=1}
\left(2q_i+u_i+d_i \right) \ , &  \cr[ {\rm SU} (2)]^2 {\rm U}(1)_X: & \ \ \ \
\ \
A_2= \sum^ 3_{i=1} \left(3q_i+l_i \right)+h_1+h_2\ , &  \cr \left[ {\rm U}
(1)_Y
\right]^2 {\rm U}(1)_X: & \ \ \ \ \ \ A_1= \sum^ 3_{i=1} \left({1 \over
3}q_i+{8 \over 3}u_i+{2 \over 3}d_i+l_i+2e_i \right)+h_1+h_2\ , &  \cr {\rm U}
(1)_Y[ {\rm U} (1)]^2_X: & \ \ \ \ \ \ A^{\prime}_ 1= \sum^ 3_{i=1}
\left[q^2_i-2u^2_i+d^2_i-l^2_i+e^2_i \right]-h^2_1+h^2_2\ . & (16) \cr} $$
The last anomaly to be considered in principle is
$ \left[ {\rm U} (1)_X \right]^3.$ It must take a precise value depending
on the $ {\rm U} (1)_X $ Kac-Moody level if the
anomalies are cancelled by the Green-Schwarz mechanism and it should vanish
if the the other anomalies vanish. Since the theory could have other
Standard Model singlets charged under $U(1)_X$ with no vev's, we don't consider
$A_X$
in this paper. For the same reason we do not consider mixed
gravitational anomalies.

It is useful to use the variable $ x $
$$ x=h_1+q_3+d_3=h_1+l_3+e_3, \ \ \ \ \ h_2+q_3+u_3=0 \eqno (17)$$
to write the following linear combinations of the anomalies
$$ \eqalignno{ A_1+A_2-2A_3 & =2 x + \sum^ 2_{a=1} \left[{2 \over 3}
\left(q_{a3}+u_{a3} \right)-{4 \over 3} \left(q_{a3}+d_{a3} \right)+2
\left(l_{a3}+e_{a3} \right) \right] \ , &  \cr A_3+3 \left(h_1+h_2 \right) & =
\sum^{ }_ a \left[ \left(q_{a3}+u_{a3} \right)+ \left(q_{a3}+d_{a3} \right)
\right]+3x \ . & (18) \cr} $$

For a given mass texture, the charge differences $ \left(q_{a3}+u_{a3}
\right), $ $ \left(q_{a3}+d_{a3} \right), $ $ \left(l_{a3}+e_{a3} \right) $
are fixed for $ a=1,2. $ and so is the r.h.s. of eq.(18). The
remaining freedom in the charge assignement can then be utilized to cancel
the
anomalies. It is appropriate to approach this issue in terms of the
familiar invariances of the SM fermion mass matrices: $ {\rm SU} (3)\times
{\rm SU} (2)\times {\rm U} (1)_Y$, the baryon number $B$, the lepton number
$ L $ and the Peccei-Quinn symmetry $ P, $ with a charge $1$ for all the
matter fields and  $-2$ for the two Higgs doublets.
Since the charge $X$ of the $U(1)_X$ has to commute with the  Standard
Model gauge group the allowed shifts in $X$ are given by a linear combination
of 4 abelian
charges:
$$ \hat X=X+\beta Y+ \gamma (B-L)+\delta (3B+L)+\zeta P \eqno (19) $$
without changing the mass textures. This redefinition can be used
to achieve an anomaly free theory by implementing either the Green-Schwarz
condition (with  $ {\rm sin}^2\theta_ W={3 \over 8}$) through the
relations, $ {3 \over 5}A_1 = A_2 = A_3$, or through the usual one,
 $A_1 = A_2 = A_3 = 0$, provided, of course, the eqs.(18) are
satisfied. In both cases one must also impose $ A'_1 = 0$.
Now, since $(B-L)$ and, obviously, $Y$ have no SM anomalies, one
can only take advantage of the shifts $\delta$ and $\zeta$ to change
$A_1,A_2,A_3.$ Then, one can use a combination of  $(B-L)$ and $Y$
to obtain  $ A'_1 = 0$. However $A'_1$ and, a fortiori,$A_1,A_2,A_3,$
are all invariant under the following shift
$$
\hat X=X+ \eta[ A_1Y-(A_1+A_2+{4\over 3}A_3-2h_1-2h_2)(B-L)] \ . \eqno(20) $$
This residual freedom
is tacitly assumed in our results from now on. It could eventually be fixed
by studying other physical consequences of the $U(1)_X$ gauge symmetry than
the mass matrices themselves.

Therefore the redefinition (19) of the charge can be used to adjust the
anomalies in (16) without affecting the combinations (18).
As we shall see in the next paragraph, those combinations have in the
model considered a direct physical meaning: they can be expressed in
terms of the determinants of the mass matrices. This is certainly an
interesting feature of the proposed solution for the Yukawa hierarchies.
It is worth remarking that the equation $A'_1 = 0$
is linear in the shift  parameters. This property
follows from $ {\rm Tr} \ YB^2= {\rm Tr} \ YL^2=0. $
\vskip 24pt

{\bf 5.}\nobreak\  We now turn to the implementation
of the anomaly cancellation in
the specific models. Let us first assume  that the charge differences
are all positive, but without restricting the values of $ h_1 $ and $ h_2. $
Then we can readily
write the relations between the Yukawa determinants and anomaly
combinations as follows
$$ \eqalignno{ {\rm det} \left(Y_UY^{-2}_DY^3_L \right) & =\varepsilon^{
3/2 \left(A_2+A_1-2A_3 \right)}\ , &  \cr {\rm det} \left(Y_UY_D \right) &
=\varepsilon^{ A_3+3 \left(h_1+h_2 \right)}\ , & (21) \cr} $$
which can also be combined into the relation
$$ {\rm det} \left(Y^{-2}_DY^2_L \right)=\varepsilon^{ A_2+A_1-{8 \over
3}A_3-2 \left(h_1+h_2 \right)}\ . \eqno (22) $$
Clearly, the anomaly combinations appearing in these mass determinants are
precisely
those that are invariant under the symmetries of the mass matrices given
by (19). The first of Eqs.(21) is independent of the value of
$ h_1+h_2. $ It is
the best-suited relation to decide if the anomalies can be put to zero
or not, using the known fermion masses.

For positive charge differences, any  mass matrices that approximately
satisfy the phenomenological mass ratios (2) entail the following results :
$$ \eqalignno{ A_1+A_2-2A_3 & \simeq 12+2x \ , &  \cr A_1+A_2-{8 \over 3}A_3 &
\simeq 2 \left(h_1+h_2 \right)\ . & (23) \cr} $$
The vanishing of the anomalies would require the physically excluded value $
x=-6. $
However, as noticed in [3] and generalized in [4], it is possible to
implement the Green-Schwarz mechanism with $ {\rm sin}^2\theta_ W={3 \over 8}
$ at the unification scale [7] if $ A_2=A_3={3 \over 5}A_1 $ and $
h_1+h_2 =0. $ But, through the redefinition (19) of $ X, $  the
parameters $ \delta $ and $\zeta $ can always be chosen
so that, $ A_2={3 \over 5}A_1 =3(6+X)$ and  $h_1+h_2 =0 ,$ as required.
In this case, the Green-Schwarz
mechanism will cancel the anomaly and $ {\rm sin}^2\theta_ W={3 \over 8}. $

Furthermore, this result immediately generalizes if there are negative charge
differences in case (a) of the preceding paragraph, where one
additional SM
gauge singlet is assumed. Indeed, the zeroes in the
mass matrices will be filled after the renormalization by the K\"ahler
metric.
However, the determinant of the K\"ahler metrics is $ O(1) $ and the
determinants
of the mass matrices are not changed by this renormalization at the leading
order in $ \varepsilon . $ The mass relations (21) and (22) are consequently
satisfied leading to the same conclusion as for positive charge differences.

In the two singlet case, if there are negative charge differences, the
relations (21) and (22) will
change. It
is easy to show that with gauge singlets with opposite charges, $ X=+1 $ and $
X=-1, $ all
the anomalies can be put to zero. Let us consider a simple example which
illustrates the general situation, with the following lepton charge
assignment:
$$ l_{13}=2\ ,\ \ \ l_{23}=3\ ,\ \ \ e_{13}=2\ ,\ \ \ e_{23}=-9\ ,\ \ \ x=2\ ,
\eqno (24) $$
corresponding to the lepton mass matrix
$$ Y_L= \varepsilon^ x \left( \matrix{ \varepsilon^ 4  &    & \varepsilon^ 3
 &    & \varepsilon^ 2 \cr   &    &    &    &   \cr \varepsilon^ 5  &    &
\varepsilon^ 2  &    & \varepsilon^ 3 \cr   &    &    &    &   \cr
\varepsilon^ 2  &    & \varepsilon^ 5  &    & 1 \cr} \right)\ . \eqno (25) $$
The quark charge differences are taken as in (9). Thus the predictions of this
model are in perfect agreement with experimental data, including the relation
$ {\rm det} \ Y_D\simeq {\rm det} \ Y_L. $ However, the relations in (21) and
(22) get modified and read
now
$$ \eqalignno{ {\rm det} \left(Y_UY^{-2}_DY^3_L \right) & = \varepsilon^{{
3 \over 2} \left(A_2+A_1-2A_3 \right)+12+6x}\ , &  \cr {\rm det}
\left(Y^{-2}_DY^2_L \right) & =\varepsilon^{ A_1+A_2-{8 \over 3}A_3-2
\left(h_1+h_2 \right)+8+4x}\ . & (26) \cr} $$

Using the values in (2) which precisely correspond to the mass matrices,
we obtain
$$ \eqalignno{ A_2+A_1-2A_3 & \simeq 4-2x\ , &  \cr A_3+3 \left(h_1+h_2
\right) & \simeq 18+3x\ . & (27) \cr} $$
The Green-Schwarz mechanism and $ {\rm sin}^2\theta_ W={3 \over 8} $ can be
obtained with the help of the relations
$$ \eqalignno{{ 3 \over 5}A_1 & =A_2=A_3=6-3x\ , &  \cr h_1+h_2  &
=4+2x\ . & (28) \cr} $$

However, it is now possible to have all anomalies
$ A_i=0 $ $ (i=1,2,3) $ if $
x=2 $ and $ h_1+h_2=8. $
Interestingly enough, this value of $ x $ gives $ \lambda_ b/\lambda_ t\simeq
\varepsilon^ 2, $ requiring $ {\rm tan} \ \beta \sim 2 $ to fit
the experimental masses.
As a matter of fact this is the unique prediction for the solutions
with vanishing anomalies, using
the most successful textures (9), (13), (14), and lepton mass matrices that
fulfill (2). More precisely, there are more assignements of charge
differences for leptons corresponding to zero anomalies, but for all
of them $x = 2$ and $h_1 + h_2 = 8$.
 Indeed, generalizing the relations (21,22) to account for
possible negative charge differences and requiring $ A_i=0 $  we get the
equation
$$ 16n_U-4(4+x)n_D+6(\mu +x)n_L=18+3x \eqno (29) $$
where $ \mu =2,4 $ correspond to $ \varepsilon^ \mu \sim \left(m_\mu /m_\tau
\right), \left(m_e/m_\tau \right), $ respectively. In (29) $ n_i=0 $ $
(i=U,D,L) $ if
all charge differences are positive in $ Y_i $ and $ n_i=1 $ if there is one
column of
negative charges in $ Y_i, $ leading to a modification of (23). A further case
to
be analyzed correspond to $ n_L=2 $ and $ \mu =3$, corresponding to a lepton
mass matrix with two columns of negative charge differences. The only solution
of
(29) is $ n_U=n_D=0, $
$ n_L=1, $ $ \mu =2 $ and $ x=2, $ which is precisely the case given by (9)
and (24)
hereabove.

All the other structures with two singlets and negative charge differences must
use the Green-Schwarz mechanism to cancel anomalies and have generally
$ h_1+h_2 \not=0.$
Therefore, if the hierarchy in the mass matrices is a consequence of an
horizontal $ U(1) $ symmetry, the vanishing of the mixed anomalies requires $
{\rm tan} \ \beta \sim 2. $
This is to be added as a possibility to the previously considered
Green-Schwarz mechanism which predicts $ {\rm sin}^2\theta_ W={3 \over 8}$.
In either case, the shifts in (19) are
instrumental, as they are also to make the $A'_1 $ anomaly  to vanish.
\vskip 24pt

{\bf 6.}\nobreak\ We now turn to specific models with definite integral charges
for the various
fields. Even if, in general, abelian charges can be rational numbers,
the string origin of the present $U(1)_X$ gauge group gives integral
charges in explicit model constructions [9].

We rewrite the anomalies as functions of the charges $
\left(q_3,u_3,d_3,l_3,e_3 \right) $
and charge differences to be taken from the allowed structures listed above.
More specifically, for the quarks we use, as before, the best textures (9),
(13) and (14) and for the leptons all the assignements compatible with the
known masses. We consider two cases:

- 1 singlet, $ {3 \over 5}A_1=A_2=A_3, $ $ A^{\prime}_ 1=0 $ (Green-Schwarz
case).

The models are characterized by the condition $ l_{13}=l_{23}=3n, $ $ n\in
 Z $ necessary in order to obtain integral charges as solutions of the
anomaly conditions. We obtain
three one-parameter solutions, displayed in the Table 2.

- 2 singlets, $ A_1=A_2=A_3=A^{\prime}_ 1=0, $ $ h_1+h_2=8, $ $ x=2. $

This case, characterized by $ l_{13}+l_{23}=3n-8, $ $ n\in Z $ has only
a one-parameter
solution with integral charges, displayed in the last column of the Table 2.
\vskip 36pt

$$ \vbox{\offinterlineskip\halign{
& \vrule#& \strut\kern.3em# \kern0pt
& \vrule#& \strut\kern.3em# \kern0pt
& \vrule#& \strut\kern.3em# \kern0pt
& \vrule#& \strut\kern.3em# \kern0pt
& \vrule#& \strut\kern.3em# \kern0pt
&\vrule#\cr
\noalign{\hrule}
height2pt& \omit&& \omit&& \omit&& \omit&& \omit&\cr
&$\displaystyle \hfill \matrix{ {\rm Number} \cr {\rm singlets} \cr}
\hfill$&&$\displaystyle \hfill 1 \hfill$&&$\displaystyle \hfill 1
\hfill$&&$\displaystyle \hfill 1 \hfill$&&$\displaystyle \hfill 2 \hfill$&\cr
height2pt& \omit&& \omit&& \omit&& \omit&& \omit&\cr
\noalign{\hrule}
height2pt& \omit&& \omit&& \omit&& \omit&& \omit&\cr
&$\displaystyle \hfill q_1 \hfill$&&$\displaystyle \hfill 3+q
\hfill$&&$\displaystyle \hfill 3+q \hfill$&&$\displaystyle \hfill 3+q
\hfill$&&$\displaystyle \hfill 3+q \hfill$&\cr
height2pt& \omit&& \omit&& \omit&& \omit&& \omit&\cr
&$\displaystyle \hfill q_2 \hfill$&&$\displaystyle \hfill 2+q
\hfill$&&$\displaystyle \hfill 2+q \hfill$&&$\displaystyle \hfill 2+q
\hfill$&&$\displaystyle \hfill 2+q \hfill$&\cr
height2pt& \omit&& \omit&& \omit&& \omit&& \omit&\cr
&$\displaystyle \hfill q_3 \hfill$&&$\displaystyle \hfill q
\hfill$&&$\displaystyle \hfill q \hfill$&&$\displaystyle \hfill q
\hfill$&&$\displaystyle \hfill q \hfill$&\cr
height2pt& \omit&& \omit&& \omit&& \omit&& \omit&\cr
&$\displaystyle \hfill u_1 \hfill$&&$\displaystyle \hfill 4+q
\hfill$&&$\displaystyle \hfill 4+q \hfill$&&$\displaystyle \hfill 4+q
\hfill$&&$\displaystyle \hfill -2-4q \hfill$&\cr
height2pt& \omit&& \omit&& \omit&& \omit&& \omit&\cr
&$\displaystyle \hfill u_2 \hfill$&&$\displaystyle \hfill 1+q
\hfill$&&$\displaystyle \hfill 1+q \hfill$&&$\displaystyle \hfill 1+q
\hfill$&&$\displaystyle \hfill -5-4q \hfill$&\cr
height2pt& \omit&& \omit&& \omit&& \omit&& \omit&\cr
&$\displaystyle \hfill u_3 \hfill$&&$\displaystyle \hfill -1+q
\hfill$&&$\displaystyle \hfill -1+q \hfill$&&$\displaystyle \hfill -1+q
\hfill$&&$\displaystyle \hfill -7-4q \hfill$&\cr
height2pt& \omit&& \omit&& \omit&& \omit&& \omit&\cr
&$\displaystyle \hfill d_1 \hfill$&&$\displaystyle \hfill 2-3q
\hfill$&&$\displaystyle \hfill 4-3q \hfill$&&$\displaystyle \hfill 4-3q
\hfill$&&$\displaystyle \hfill 2+2q \hfill$&\cr
height2pt& \omit&& \omit&& \omit&& \omit&& \omit&\cr
&$\displaystyle \hfill d_2 \hfill$&&$\displaystyle \hfill 1-3q
\hfill$&&$\displaystyle \hfill 3-3q \hfill$&&$\displaystyle \hfill 3-3q
\hfill$&&$\displaystyle \hfill 1+2q \hfill$&\cr
height2pt& \omit&& \omit&& \omit&& \omit&& \omit&\cr
&$\displaystyle \hfill d_3 \hfill$&&$\displaystyle \hfill 1-3q
\hfill$&&$\displaystyle \hfill 3-3q \hfill$&&$\displaystyle \hfill 3-3q
\hfill$&&$\displaystyle \hfill 1+2q \hfill$&\cr
height2pt& \omit&& \omit&& \omit&& \omit&& \omit&\cr
&$\displaystyle \hfill l_1 \hfill$&&$\displaystyle \hfill 1-3q
\hfill$&&$\displaystyle \hfill 4-3q \hfill$&&$\displaystyle \hfill 2-3q
\hfill$&&$\displaystyle \hfill -8-3q \hfill$&\cr
height2pt& \omit&& \omit&& \omit&& \omit&& \omit&\cr
&$\displaystyle \hfill l_2 \hfill$&&$\displaystyle \hfill 1-3q
\hfill$&&$\displaystyle \hfill 2-3q \hfill$&&$\displaystyle \hfill 6-3q
\hfill$&&$\displaystyle \hfill -6-3q \hfill$&\cr
height2pt& \omit&& \omit&& \omit&& \omit&& \omit&\cr
&$\displaystyle \hfill l_3 \hfill$&&$\displaystyle \hfill 1-3q
\hfill$&&$\displaystyle \hfill 3-3q \hfill$&&$\displaystyle \hfill 1-3q
\hfill$&&$\displaystyle \hfill -9-3q \hfill$&\cr
height2pt& \omit&& \omit&& \omit&& \omit&& \omit&\cr
&$\displaystyle \hfill e_1 \hfill$&&$\displaystyle \hfill 4+q
\hfill$&&$\displaystyle \hfill 3+q \hfill$&&$\displaystyle \hfill -1+q
\hfill$&&$\displaystyle \hfill 13+6q \hfill$&\cr
height2pt& \omit&& \omit&& \omit&& \omit&& \omit&\cr
&$\displaystyle \hfill e_2 \hfill$&&$\displaystyle \hfill 2+q
\hfill$&&$\displaystyle \hfill 3+q \hfill$&&$\displaystyle \hfill 5+q
\hfill$&&$\displaystyle \hfill 1+6q \hfill$&\cr
height2pt& \omit&& \omit&& \omit&& \omit&& \omit&\cr
&$\displaystyle \hfill e_3 \hfill$&&$\displaystyle \hfill q
\hfill$&&$\displaystyle \hfill q \hfill$&&$\displaystyle \hfill 2+q
\hfill$&&$\displaystyle \hfill 10+6q \hfill$&\cr
height2pt& \omit&& \omit&& \omit&& \omit&& \omit&\cr
&$\displaystyle \hfill x \hfill$&&$\displaystyle \hfill 0
\hfill$&&$\displaystyle \hfill 2 \hfill$&&$\displaystyle \hfill 2
\hfill$&&$\displaystyle \hfill 2 \hfill$&\cr
height2pt& \omit&& \omit&& \omit&& \omit&& \omit&\cr
&$\displaystyle \hfill \matrix{ {\rm Anomaly} \cr {\rm cancellation} \cr}
\hfill$&&$\displaystyle \hfill {\rm GS} \hfill$&&$\displaystyle \hfill {\rm
GS} \hfill$&&$\displaystyle \hfill {\rm GS} \hfill$&&$\displaystyle \hfill
{\rm Yes} \hfill$&\cr
height2pt& \omit&& \omit&& \omit&& \omit&& \omit&\cr
\noalign{\hrule}}}
 $$
\vskip 24pt
{\bf TABLE 2:}
The models with integral charges corresponding to one singlet (Green-Schwarz)
and two singlets (zero anomalies). The parameter $ x $ is defined as $ {\rm
tg} \ \beta ={m_t \over m_b} \lambda ^x $ and
 $ q $ is an integer. Only the solutions (9), (13) and (14) were used for the
table.

\vskip 36pt

{\bf 7.} To conclude, we analyzed the connection between the mass matrices
and the anomaly cancellation conditions for a horizontal $U(1)_X$ gauge
symmetry, spontaneously broken by one or two Standard Model gauge singlets
of opposite $X$ charges. We first classified the
acceptable mass matrices according to the known data. We remarked that the
renormalization coming from the K\"ahler potential is generally important
if the mass matrices initially had some zeroes. The anomaly analysis reveals
that with two singlets and negative charge differences for the leptons
there is a unique
solution with integral charges and zero anomalies which predicts
$tg \beta \sim {m_t \over m_b} \lambda^2$. Restricting to the one singlet
case, corresponding to the solution (9) there are three solutions with integral
charges corresponding to the
Green-Schwarz mechanism. In this case either $tg \beta \sim {m_t \over m_b}$
or $tg \beta \sim {m_t \over m_b} \lambda^2$.
\vskip 12pt
{\bf Acknowledgments:}
One of us (E.D.) would like to thank Pierre Bin\'etruy for interesting
discussions and useful comments.

\vfill\eject
\vfill\eject
\centerline{{\bf APPENDIX}}
\vskip 24pt
We display here the explicit expressions of the diagonalyzing angles for the
mass matrices defined in eqs.(1) and (4), necessary in order to compute
$V_{CKM}$.
In the $ Y_{33}\geq Y_{ij} $ hypothesis we find the following results, using
the
parametrization (4):

\parindent=0truecm

$ \matrix{ S_{13}\simeq \hfill \cr{ Y_{13}Y^3_{33}+
\left(Y_{11}Y_{31}+Y_{12}Y_{32} \right)Y^2_{33}+
\left(Y_{12}Y_{23}-Y_{13}Y_{22} \right)Y_{22}Y_{33}-
\left(Y_{13}Y_{21}-Y_{11}Y_{23} \right)Y_{21}Y_{33}+
\left(Y_{11}Y_{22}-Y_{12}Y_{21} \right) \left(Y_{21}Y_{32}-Y_{31}Y_{22}
\right) \over Y^4_{33}- \left(Y^2_{11}+Y^2_{12}+Y^2_{21}+Y^2_{22}
\right)Y^2_{33}}\ , \hfill \cr} $

$ \matrix{ S_{23}\simeq \hfill \cr{ Y_{23}Y^3_{33}+
\left(Y_{22}Y_{32}+Y_{21}Y_{31} \right)Y^2_{33}+
\left(Y_{21}Y_{13}-Y_{23}Y_{11} \right)Y_{11}Y_{33}-
\left(Y_{23}Y_{12}-Y_{22}Y_{13} \right)Y_{12}Y_{33}+
\left(Y_{11}Y_{22}-Y_{12}Y_{21} \right) \left(Y_{12}Y_{31}-Y_{32}Y_{11}
\right) \over Y^4_{33}- \left(Y^2_{11}+Y^2_{12}+Y^2_{21}+Y^2_{22}
\right)Y^2_{33}}\ , \hfill \cr} $
\parindent=20pt
$$ \eqalignno{ S^{\prime}_{ 13} & ={Y_{12}Y_{23}-Y_{13}Y_{22}+
\left(S_{13}Y_{22}-S_{23}Y_{12} \right)Y_{33} \over
Y_{11}Y_{22}-Y_{12}Y_{21}}\ , &  \cr S^{\prime}_{ 23} &
={Y_{13}Y_{21}-Y_{11}Y_{23}- \left(S_{13}Y_{21}-S_{23}Y_{11} \right)Y_{33}
\over Y_{11}Y_{22}-Y_{12}Y_{21}}\ . & (A1) \cr} $$
The general expressions for $ S_{12} $ and $ S^{\prime}_{ 12} $ are more
involved. However, if one of
the following conditions holds:
$$ Y_{12}Y_{32}\leq Y_{23}Y_{33}\ ,\ \ \ Y_{12}Y_{32}\leq Y_{13}Y_{33}\ ,\ \ \
Y_{12}\leq Y_{22}\ , \eqno (A2) $$
and defining, following Hall-Ra$ \check {\rm s} $in
$$ \eqalignno{\tilde Y_{22} & =Y_{22}-S^{\prime}_{
23}Y_{23}-S_{23}Y_{32}+S_{23}S^{\prime}_{ 23}Y_{33}\simeq Y_{22}-Y_{23}Y_{32}
&  \cr\tilde Y_{11} & =Y_{11}-S^{\prime}_{
13}Y_{13}-S_{13}Y_{31}-S_{13}S_{23}Y_{21}+S_{13}S^{\prime}_{ 13}Y_{33}\simeq
Y_{11}-Y_{13}Y_{31}-Y_{13}Y_{23}Y_{21} &  \cr\tilde Y_{12} &
=Y_{12}-S^{\prime}_{
23}Y_{13}-S_{13}S_{23}Y_{22}-S_{13}Y_{32}+S_{13}S^{\prime}_{ 23}Y_{33}\simeq
Y_{12}-Y_{13}Y_{32}-Y_{13}Y_{23}Y_{22} &  \cr\tilde Y_{21} &
=Y_{21}-S^{\prime}_{ 13}S^{\prime}_{ 23}Y_{22}-S^{\prime}_{
13}Y_{23}-S_{23}Y_{31}+S_{23}S^{\prime}_{ 13}Y_{33}\simeq
Y_{21}-Y_{23}Y_{31}-Y_{22}Y_{31}Y_{32}\ , &  \cr  &   &  (A3) \cr} $$
we obtain the approximate relations
$$ \eqalignno{ S_{12} & \simeq{\tilde Y_{12}\tilde Y_{22}+\tilde Y_{11}\tilde
Y_{21} \over\tilde Y^2_{22}-\tilde Y^2_{11}+\tilde Y^2_{12}+\tilde
Y^2_{21}}\simeq{\tilde Y_{12} \over\tilde Y_{22}}+{\tilde Y_{11}\tilde Y_{21}
\over\tilde Y^2_{22}} &  \cr S^{\prime}_{ 12} & \simeq{\tilde Y_{21}+\tilde
Y_{11}S_{12} \over\tilde Y_{22}+\tilde Y_{12}S_{12}}\simeq{\tilde Y_{21}
\over\tilde Y_{22}}+{\tilde Y_{11}\tilde Y_{12} \over\tilde Y^2_{22}}\ . &
(A4) \cr} $$
\vfill\eject
\centerline{{\bf REFERENCES}}

\vskip 24pt

\item{$\lbrack$1$\rbrack$} C.D. Froggatt and H.B. Nielsen, Nucl. Phys. B
{\bf 147} (1979) 277.
\item{\nobreak\ \nobreak\ \nobreak\ \nobreak\ } S. Dimopoulos, Phys. Lett. B
{\bf 129} (1983) 417;
\item{\nobreak\ \nobreak\ \nobreak\ \nobreak\ } J. Bagger, S. Dimopoulos, E.
Masso and M. Reno, Nucl. Phys. B {\bf 258 }
(1985) 565;
\item{\nobreak\ \nobreak\ \nobreak\ \nobreak\ } Z.G. Berezhiani, Phys. Lett. B
{\bf 129} (1983) 99; B{\bf\ 150} (1985) 177.
\item{\nobreak\ \nobreak\ \nobreak\ \nobreak\ \nobreak\ } J. Bijnens and
C. Wetterich, Nucl. Phys. B {\bf 283} (1987) 237.

\item{$\lbrack$2$\rbrack$} M. Leurer, Y. Nir and N. Seiberg, Nucl. Phys. B
{\bf 398} (1993) 319, Nucl.
Phys. B {\bf 420} (1994) 468;
\item{\nobreak\ \nobreak\ \nobreak\ \nobreak\ \nobreak\ } Y. Nir and N.
Seiberg, Phys. Lett. B {\bf 309} (1993) 337.

\item{$\lbrack$3$\rbrack$} L.E. Ib\`a$ \tilde {\rm n} $ez and G.G. Ross,
Phys. Lett. B {\bf 332} (1994) 100.

\item{$\lbrack$4$\rbrack$} P. Bin\'etruy and P. Ramond, LPTHE 94/115,
hep-ph/9412385.

\item{$\lbrack$5$\rbrack$} P. Ramond, R.G. Roberts and G.G. Ross, Nucl.
Phys. B {\bf 406} (1993) 19.

\item{$\lbrack$6$\rbrack$} V. Jain and R. Shrock, ITP-SB-94-55.
\item{\nobreak\ \nobreak\ \nobreak\ \nobreak\ } E. Papageorgiu, LPTHE Orsay
40/94.

\item{$\lbrack$7$\rbrack$} L.E. Ib\`a$ \tilde {\rm n} $ez, Phys. Lett. B
{\bf 303} (1993) 55.

\item{$\lbrack$8$\rbrack$} M. Dine, N. Seiberg and E. Witten, Nucl. Phys. B
{\bf 289} (1987) 585;
\item{\nobreak\ \nobreak\ \nobreak\ \nobreak\ \nobreak\ } J. Atick, L. Dixon
and A. Sen, Nucl. Phys. B {\bf 292} (1987) 109;
\item{\nobreak\ \nobreak\ \nobreak\ \nobreak\ \nobreak\ } M. Dine, I.
Ich\'enoise and N. Seiberg, Nucl. Phys. B {\bf 293} (1987)
253.

\item{$\lbrack$9$\rbrack$} A. Font, L.E. Ib\`a$ \tilde {\rm n} $ez, H.P.
Nilles and F. Quevedo, Nucl. Phys. B
{\bf 307} (1988) 109; Phys. Lett. B {\bf 210} (1988) 101;
\item{\nobreak\ \nobreak\ \nobreak\ \nobreak\ \nobreak\ \nobreak\ } J.A.
Casas, E.K. Katehou and C. Mu$ \tilde {\rm n} $oz, Nucl. Phys. B {\bf 317}
(1989)
171;
\item{\nobreak\ \nobreak\ \nobreak\ \nobreak\ \nobreak\ \nobreak\ } J.A. Casas
and C. Mu$ \tilde {\rm n} $oz, Phys. Lett. B {\bf 209} (1988) 214; Phys.
Lett. B {\bf 214} (1988) 63;
\item{\nobreak\ \nobreak\ \nobreak\ \nobreak\ \nobreak\ \nobreak\ } A. Font,
L.E. Iba$ \tilde {\rm n} $ez, F. Quevedo and A. Sierra, Nucl. Phys. B
{\bf 331} (1990) 421.

\item{$\lbrack$10$\rbrack$} M. Olechowski and S. Pokorski, Phys. Lett. B {\bf
257}
(1991) 388.

\item{$\lbrack$11$\rbrack$} L.J. Hall and A. Ra$ \check {\rm s} $in, Phys.
Lett. B {\bf 315} (1993) 164.

\item{$\lbrack$12$\rbrack$} S. Weinberg, in \lq\lq A Festschrift for I.I.
Rabi\rq\rq\ (Trans. N.Y. Acad.
Sci., Ser.II (1977), v.38), p.185;
\item{\nobreak\ \nobreak\ \nobreak\ } F. Wilczek and A. Zee, Phys. Lett. B
{\bf 70} (1977) 418;
\item{\nobreak\ \nobreak\ \nobreak\ } T. Maehara and T. Yanagida, Prog. Theor.
Phys. {\bf 60} (1978) 822;
\item{\nobreak\ \nobreak\ \nobreak\ } F. Wilczek and A. Zee, Phys. Rev. Lett.
{\bf 42} (1979) 421.

\item{$\lbrack$13$\rbrack$} H. Fritsch, Phys. Lett. B {\bf 70} (1977)
436; Phys. Lett. B {\bf 73}
(1978) 317;
\item{\nobreak\ \nobreak\ \nobreak\ \nobreak\ \nobreak\ \nobreak\ \nobreak\
\nobreak\ \nobreak\ } F.J. Gilman and Y. Nir, Ann. Rev. Nucl. Part. Sci. {\bf
40} (1990)
213;
\item{\nobreak\ \nobreak\ \nobreak\ \nobreak\ \nobreak\ \nobreak\
\nobreak\ } P. Kaus and S. Meshkov, Mod. Phys. Lett. A {\bf 3} (1988)
1251.

\item{$\lbrack$14$\rbrack$} J. Harvey, P. Ramond and D. Reiss, Phys. Lett. B
{\bf 92} (1980) 309;
\item{\nobreak\ \nobreak\ \nobreak\ \nobreak\ \nobreak\ } S. Dimopoulos, L.J.
Hall and S. Raby, Phys. Rev. Lett. {\bf 68} (1992)
1984; Phys. Rev. D {\bf 45} (1992) 4195;
\item{\nobreak\ \nobreak\ \nobreak\ \nobreak\ \nobreak\ } H. Arason, D.J.
Casta$ \tilde {\rm n} $o, J. Ramond and E.J. Piard, Phys. Rev. D
{\bf 47} (1993) 232;
\item{\nobreak\ \nobreak\ \nobreak\ \nobreak\ \nobreak\ } G.F. Giudice, Mod.
Phys. Lett. A {\bf 7} (1992) 2429.

\item{$\lbrack$15$\rbrack$} M.S. Chanowitz, J. Ellis and
M.K. Gaillard, Nucl. Phys. B {\bf 128}
(1977) 506;
\item{\nobreak\ \nobreak\ \nobreak\ \nobreak\ } A. Buras, J. Ellis, M.K.
Gaillard and D.V. Nanopoulos, Nucl. Phys.
B {\bf 135} (1978) 66;
\item{\nobreak\ \nobreak\ \nobreak\ \nobreak\ } H. Arason, D.J. Casta$ \tilde
{\rm n} $o, B. Keszthelyi, S. Mikaelian, E.J. Piard,
P. Ramond and B.D. Wright, Phys. Rev. Lett. {\bf 67} (1991) 2933.

\item{$\lbrack$16$\rbrack$} Y. Grossman and Y. Nir, hep-ph/9502418.

\item{$\lbrack$17$\rbrack$} E. Dudas, S. Pokorski and C.A. Savoy, work in
progress.

\item{$\lbrack$18$\rbrack$} P. Bin\'etruy and E. Dudas, LPTHE 95/18, SPhT
T95/042.
\end